\input harvmac
\newcount\figno
\figno=0
\def\fig#1#2#3{
\par\begingroup\parindent=0pt\leftskip=1cm\rightskip=1cm\parindent=0pt
\baselineskip=11pt
\global\advance\figno by 1
\midinsert
\epsfxsize=#3
\centerline{\epsfbox{#2}}
\vskip 12pt
{\bf Fig. \the\figno:} #1\par
\endinsert\endgroup\par
}
\def\figlabel#1{\xdef#1{\the\figno}}
\def\encadremath#1{\vbox{\hrule\hbox{\vrule\kern8pt\vbox{\kern8pt
\hbox{$\displaystyle #1$}\kern8pt}
\kern8pt\vrule}\hrule}}

\overfullrule=0pt

%
\def\tilde{\widetilde}

\def\Z{{\bf Z}}
\def\T{{\bf T}}
\def\S{{\bf S}}
\def\R{{\bf R}}

\font\zfont = cmss10 

\def\bigone{\hbox{1\kern -.23em {\rm l}}}
\def\ZZ{\hbox{\zfont Z\kern-.4emZ}}

\Title{hep-th/9512219, IASSNS-HEP-96-01}
{\vbox{\centerline{Five-branes and $M$-Theory On An Orbifold}}}
\smallskip
\centerline{Edward Witten\foot{Research supported in part
by NSF  Grant PHY92-45317.}}
\smallskip
\centerline{\it School of Natural Sciences, Institute for Advanced Study}
\centerline{\it Olden Lane, Princeton, NJ 08540, USA}\bigskip

\medskip

\noindent
We relate Type IIB superstrings compactified to six dimensions on K3 to
an eleven-dimensional theory compactified on $({\bf S}^1)^5/\Z_2$.
Eleven-dimensional five-branes enter the story in an interesting
way.

\Date{January, 1996}

\newsec{Introduction}
By now, there is substantial evidence for the existence of an
eleven-dimensional quantum theory with eleven-dimensional supergravity
as its long wave-length limit.  Moreover, the theory contains
two-branes and five-branes at least macroscopically, and some of their
properties are known; for instance, the $\kappa$-invariant 
Bergshoeff-Sezgin-Townsend action\ref\berg{E. Bergshoeff, E. Sezgin, 
and P. K. Townsend, ``Supermembranes And Eleven-Dimensional
Supergravity,'' Phys. Lett. {\bf 189B} (1987) 75.}
 describes the long wavelength
excitations of a macroscopic two-brane.

The description by eleven-dimensional supergravity with two-branes and
five-branes is expected to be valid when all characteristic length scales
(of a space-time and the branes that it contains) are large compared
to the Planck length.  One also has some information about the behavior
under certain conditions when some dimensions of space-time are 
{\it small} compared to the Planck scale.  For instance, the
eleven-dimensional 
``$M$-theory'' (where $M$ stands for magic, mystery, or membrane, according
to taste) on $X\times \S^{1}$, with $X $ any ten-manifold,
is equivalent to Type IIA on $X$, with a Type IIA string coupling constant
that becomes  small when the radius of $\S^1$ goes to zero.  Likewise,
the $M$-theory on $Y\times {\rm K3}$, with $Y$ a seven-manifold, is
equivalent to the heterotic string on $Y\times \T^3$, and the $M$-theory
on $X\times \S^1/Z_2$, with $X$ a ten-manifold, is equivalent to the 
$E_8\times E_8$ heterotic string on $X$; in each case, the string
coupling constant becomes small when the volume of the last factor goes
to zero.   

The evidence for the existence of the $M$-theory (beyond the
consistency of the classical low energy theory) 
comes mainly from the success of statements deduced from the relations
 of the $M$-theory to strings.   Even a few more similar examples
might therefore significantly enrich the story.
The purpose of the present paper is to add one more such example,
 by arguing that the $M$-theory on $Z\times (\S^1)^5/\Z_2$
is equivalent to the Type IIB superstring on $Z\times {\rm K3}$.  Here
$Z$ is an arbitrary six-manifold, but as usual in such arguments, by
scaling up the metric of $Z$, one can reduce to the case that $Z=\R^6$.
In fact, once an equivalence is established between the $M$-theory on $Z\times
(\S^1)^5/\Z_2$ and Type IIB on $Z\times {\rm K3}$ when $Z$ is large, it can
be followed into the region of small $Z$.

The equivalence of the $M$-theory on $\R^6\times (\S^1)^5/\Z_2$ with
Type IIB on $\R^6\times {\rm K3}$ was also conjectured recently
by Dasgupta and Mukhi \ref\mukhi{K. Dasgupta and S. Mukhi,
``Orbifolds oF $M$-Theory,'' hepth/9512196.}
who independently pointed out a problem -- involving
anomaly cancellation and the distribution of the twisted sectors among
fixed points -- that will be addressed below.  Some general
comments about Type IIB on K3 as an $M$-theory orbifold
were also made recently
by Hull \ref\hull{C. M. Hull.   ``String Dynamics at Strong Coupling,''
hepth/9512181.}.

\newsec{\it The Low Energy Supergravity}

Compactification of the Type IIB superstring on ${\rm K3}$ gives a
six-dimensional theory with a chiral supersymmetry which (upon
toroidal compactification to four dimensions) is related to $N=4$
supersymmetry in $D=4$.  We will  call this six-dimensional
chiral $N=4$ supersymmetry  (though the number of supercharges is only
twice the minimum possible number in $D=6$).  

The supergravity multiplet of chiral $N=4$ supergravity contains,
in addition to the graviton, five self-dual tensors (that is two-forms
with self-dual field strength) plus gravitinos.  The graviton in six
dimensions has nine helicity states, while the self-dual tensor has three,
so the total number of bosonic helicity states is $9+5\cdot 3=24$; the
gravitinos likewise have 24 helicity states.  The supergravity multiplet
has gravitational anomalies (which cannot be canceled by the Green-Schwarz
mechanism alone), so any consistent theory with chiral $N=4$ supergravity
in six dimensions must contain matter multiplets also.

There is actually 
only one possible matter multiplet in chiral
$N=4$ supersymmetry.  It is the tensor multiplet, which contains
five spin zero bosons, an anti-self-dual antisymmetric tensor
(that is a two-form field whose field strength is anti-self-dual)
with three helicity states, and $5+3=8$ helicity states of chiral
fermions.  Cancellation of gravitational anomalies requires that the number
of tensor multiplets be precisely 21.  

Using only the low energy supergravity, one can deduce (for a survey
of such matters
see \ref\salam{A. Salam and E. Sezgin, ``Supergravities In Diverse
Dimensions,'' North-Holland/World-Scientific (1989).}) that the
moduli space ${\cal M}$ 
of vacua is locally the homogeneous space $SO(21,5)/SO(21)
\times SO(5)$.  In the particular case of a chiral $N=4$ theory obtained
by compactification of Type IIB on K3, the global structure 
is actually (as asserted in equation (4.16)  
of \ref\witten{E. Witten,
``String Theory Dynamics In Various Dimensions,'' Nucl. Phys. {\bf B443}
(1995) 85.}; see
\ref\otherasp{P. Aspinwall and D. Morrison, ``$U$-Duality And
Integral Structures,'' hepth/9505025.} for a more precise justification)
${\cal M}=SO(21,5;{\bf Z})\backslash SO(21,5)/SO(21)\times SO(5)$.
This depends on knowledge of conformal field theory $T$-duality on K3
\ref\aspinwall{P. Aspinwall and D. Morrison, ``String Theory On
K3 Surfaces,'' hepth/9404151.} 
together with the $SL(2,\Z)$
symmetry of ten-dimensional Type IIB superstring theory.

Note that since there is no scalar in the chiral $N=4$ supergravity multiplet,
the dilaton is one of the $5\times 21=105$ scalars that come from the
tensor multiplets.  The $SO(21,5;{\bf Z})$ discrete symmetry mixes
up the dilaton with the other 104 scalars, relating some but not all
of the ``strong coupling'' regimes to regions of weak coupling or large
volume.

\subsec{Five-branes And The Tensor Multiplet Anomaly}

We will need some background about fivebranes and gravitational anomalies.

We want to consider a certain model of {\it global} chiral $N=4$ supergravity
with the tensor multiplet.  To do this, we begin in eleven-dimensional
Minkowski space, with coordinates $x^1,\dots, x^{11}$ ($x^1$ being the
time), and gamma matrices $\Gamma^1,\dots,\Gamma^{11}$ which obey
\eqn\jumob{\Gamma^1\Gamma^2\dots \Gamma^{11}=1.}
Now we introduce a five-brane with world-volume given by the equations
\eqn\umob{x^7=\dots = x^{11}=0.}
The presence of this five-brane breaks 
 half of the 32 space-time supersymmetries.  The 16 surviving supersymmetries
are those that obey $\Gamma^7\dots \Gamma^{11}=1$, or equivalently, in
view of \jumob, $\Gamma^1\dots \Gamma^6=1$.  Thus, the surviving 
supersymmetries are chiral in the six-dimensional sense; the world-volume
theory of the five-brane has chiral $N=4$ supersymmetry.  This is {\it global}
supersymmetry since -- as the graviton propagates in bulk -- there is
no massless graviton on the five-brane world-volume. 

Therefore, the massless world-volume fields must make up a certain number
of tensor multiplets, this being the only matter multiplet allowed by
chiral $N=4$ supergravity.  In fact, there is precisely one tensor multiplet.
The five massless scalars are simply the fluctuations in $x^7,\dots,x^{11}$;  
the massless world-volume fermions are the Goldstone fermions associated
with the supersymmetries under which the five-brane is not invariant;
and the anti-self-dual tensor has an origin that was described semiclassically
in \ref\kaplan{D. Kaplan and J. Michelson,
``Zero Modes For The $D=11$ Membrane And Five-Brane,'' hepth/9510053.}.
The assertion that the massless world-volume
excitations of the five-brane consist of precisely one tensor multiplet
can also be checked by  compactifying the $x^{11}$ direction on a circle,
and comparing to the structure of a Dirichlet four-brane of Type IIA
\ref\polch{J. Polchinski, ``Dirichlet-Branes And Ramond-Ramond
Charges,'' hepth/9510017.}.  (In compactifying the $M$-theory
to Type IIA, the five-brane wrapped around $x^{11}$ turns into a four-brane;
the tensor multiplet of $5+1$ dimensions reduces to a vector multiplet
in $4+1$ dimensions, which is the massless world-volume structure of
the Dirichlet four-brane.)

Now we want to allow fluctuations in the position of the five-brane
and compute the quantum behavior at long wavelengths.   At once we run
into the fact that the tensor multiplet on the five-brane world-volume
has a gravitational anomaly.  Without picking a coordinate system on
the five-brane world-volume, how can one cancel the anomaly in the one
loop effective action of the massless world-volume fields (even at very
long wavelengths where the one loop calculation is valid)?  

This question was first discussed by Duff, Liu, and Minasian 
\ref\duff{M. J. Duff, J. T. Liu, and R. Minasian, ``Eleven Dimensional
Origin Of String/String Duality: A One Loop Test,'' hepth/9506126.
}; what follows is a 
sort of dual version of their resolution
of the problem.\foot{In the very similar
case of ten-dimensional Type IIA five-branes, 
the dual version was worked out in unpublished work by J. Blum and J. A.
Harvey.}
The tensor multiplet anomaly {\it cannot} be cancelled,
as one might have hoped, by a world-volume Green-Schwarz mechanism.
Instead one has to cancel a world-volume effect against a bulk contribution
from the eleven-dimensional world, rather as in 
\ref\callan{C. G. Callan, Jr. and J. A. Harvey,
``Anomalies And Fermion Zero Modes On Strings And Domain Walls,''
Nucl. Phys. {\bf B250} (1985) 427 for a treatment of anomaly
cancellation for certain four-dimensional extended objects by current
inflow from the bulk.}.

This theory has in the long-wavelength description a four-form $F$ that
is closed in the absence of five-branes, but which in the presence
of five-branes obeys
\eqn\inko{dF=\delta_V}
where $\delta_V$ is a delta-function supported on the five-form world-volume
$V$.  There is here a key point in the terminology: given
a codimension $n$ submanifold $W$ of
space-time,  the symbol $\delta_W$ will  denote not really a delta
``function'' but  a closed $n$-form
supported on $W$ which integrates to one in the directions normal to $W$.
For instance, in one dimension, if $P$ is the origin on the $x$-axis,
then  $\delta_P=
\delta(x) \, dx$ where $\delta(x)$ is the ``Dirac delta
function'' and $\delta(x)\, dx$ is, therefore, a closed one-form that vanishes
away from the origin and whose integral over the $x$-axis (that is, the
directions normal to $P$) is 1.  With $\delta_V$ thus understood
as a closed five-form in the five-brane case, \inko\ is compatible with
the Bianchi identity $d(dF)=0$ and is, in fact, sometimes taken as a defining
property of the five-brane as it asserts that the five-brane couples
magnetically to $F$.

Now suppose that in the low energy expansion of the effective 
eleven-dimensional theory on a space-time $M$ there is a term
\eqn\jimoc{\Delta L= \int_M F\wedge I_7}
where $I_7$ is a gravitational Chern-Simons seven-form.  
Exactly which Chern-Simons seven-form it should be will soon become clear.
Under an infinitesimal  
diffeomorphism $x^I\to x^I+\epsilon v^I$ ($\epsilon$ being an infinitesimal
parameter and $v$ a vector field), 
$I_7$ does not transform as a tensor, but
rather $I_7\to I_7+d J_6$, where $J_6$ is a certain six-form (which depends
upon $v$).
The transformation of $\Delta L$ under a diffeomorphism is therefore
\eqn\himoc{\Delta L\to \Delta L +\int_M F\wedge dJ_6
=\Delta L-\int_M dF\wedge J_6.}
Thus, $\Delta L$ is generally covariant in the absence of five-branes,
but in the presence of a five-brane, according to \inko,
one gets
\eqn\gimoc{\Delta L\to \Delta L-\int_V J_6.}

But gravitational anomalies in $n$ dimensions involve precisely expressions
$\int J_n$ where $J_n$ is as above
(that is, $J_n$ appears in the transformation law of a Chern-Simons
$n+1$-form $I_{n+1}$ by $I_{n+1}\to I_{n+1}+dJ_n$; see 
\ref\ginsparg{L. Alvarez-Gaum\'e and P. Ginsparg,
``The Structure Of Gauge And Gravitational Anomalies,''
Ann. Phys. {\bf 161} (1985) 423.}  for an
introduction to such matters.)  Thus with the correct choice of $I_7$,
the anomaly of  $\Delta L$ in the
presence of a five-brane precisely cancels the world-volume anomaly of
the tensor multiplet.  
This is thus a case in which an interaction in the bulk is needed
to cancel on anomaly on the world-volume.
Moreover (as explained in a dual language in \duff),
the presence in eleven dimensions of the interaction $\Delta L$ can be checked
by noting that upon compactification on a circle, this interaction reduces
to the $H\wedge I_7$ term found in \ref\vafawitten{C. Vafa and E. Witten,
``A One Loop Test Of String Duality,'' hepth/9505053.}
for Type IIA superstrings; here $H$ is the usual three-form field strength
of the Type IIA theory.

What has been said to this point is sufficient for our purposes.
However, I cannot resist a further comment that involves somewhat similar
ideas.  The seven-form $F'$ dual to $F$ does not obey $dF'=0$ even in the
absence of five-branes; from the eleven-dimensional supergravity one finds
instead
\eqn\ujjo{dF'+{1\over 2}F\wedge F = 0.}
One may ask how this is compatible with the Bianchi identity $d(dF')=0$
once -- in the presence of five-branes -- one encounters a situation
with $dF\not= 0$.  The answer involves the anti-self-dual three-form
field strength $T$ on the five-brane world-volume.  According to 
equation (3.3) of \ref\newtownsend{P. Townsend, ``$D$-Branes From
$M$-Branes,'' hepth/95012062.}, this field obeys
not -- as one might expect -- $dT=0$, but rather $dT=F$.  If then in the
presence of a five-brane, \ujjo\ becomes
\eqn\gujjo{dF' +{1\over 2} F\wedge F -T \wedge \delta_V=0,}
then the Bianchi identity still works even in the presence of the five-brane.
The $T\wedge \delta_V$ term in fact follows from the coupling in
equation (3.3) of \newtownsend, which gives
a fivebrane contribution to the equation of motion of the
three-form $A$.  Thus, we get a new derivation of the relevant coupling
 and in particular of the fact that $dT=F$.

\newsec{\it Type IIB On K3}

We now come to the main focus of this paper.
One would like to understand the ``strong coupling behavior'' of the Type IIB
theory compactified on K3, 
or more precisely, the behavior as one goes to infinity in
the moduli space ${\cal M}$ of vacua.  As explained above, this theory
has  a $SO(21,5;{\bf Z})$ discrete symmetry, which 
gives many identifications of strong coupling or small volume with weak
coupling or large volume, but there remain (as in section three of \witten\
or \otherasp) 
inequivalent limits in which one can go to infinity in ${\cal M}$. 

Any limit can be reached by starting at a given point $P\in {\cal M}$ 
and then considering the one-parameter family of vacua $P_t=e^{tx}P$
where $x$ is a generator of $SO(21,5)$ and $t$ is a positive real number.
As $t\to\infty$, one approaches infinity in ${\cal M}$ in a direction
that depends upon $x$.  
In any such limit, by looking at the lightest states, one aims to find
a description by an effective ten-dimensional string theory or 
eleven-dimensional field theory.  The duality group visible (though
mostly spontaneously broken, depending on the precise choice of $P$) in this
effective theory will include  
the subgroup $\Gamma$ of $SO(21,5;\Z)$ that commutes
with $x$ (and so preserves the particular direction in which one has gone
to infinity).

As in \refs{\witten,\otherasp}, 
one really only needs to consider $x$'s that lead to a maximal set of
light states, and 
because of
the discrete $SO(21,5;{\bf Z})$, there are only finitely many cases
to consider.  We will focus here on the one limit that seems to be related
most directly to the $M$-theory.

Consider a subgroup $SO(16)\times SO(5,5) $ of $SO(21,5)$.  Let $x$ be
a generator of $SO(5,5)$ that commutes with an $SL(5)$ subgroup.
Then the subgroup of $SO(21,5;\Z)$ that commutes with $x$ -- and so
is visible if one goes to infinity in the direction determined by $x$ --
contains $SO(16)\times SL(5,\Z)$.  

Since it will play a role later, let us discuss just how $SL(5,\Z)$ can
be observed as a symmetry at infinity.  Instead of making mathematical
arguments, we will discuss another (not unrelated, as we will see) 
physical problem 
with $SO(21,5;\Z)$ symmetry, namely the compactification on a five-torus of
the $SO(32)$ heterotic string to five dimensions, with $SO(21,5;\Z)$
as the $T$-duality group.  The region at infinity in moduli space
in which there is a visible $SL(5,\Z)$ symmetry is simply the large volume
limit, with the torus large in all directions.  In what sense can
$SL(5,\Z)$ be ``observed''?  It is spontaneously broken (to a finite subgroup,
generically trivial) by the choice
of a metric on the five-torus, but, if one is free to move around
in the moduli space of large volume metrics (remaining at infinity
in ${\cal M}$) one can see that there is a spontaneously
broken $SL(5,\Z)$.

Now, actually, the relevant region at 
infinity in moduli space is parametrized by a large
metric on the torus, a $B$-field, and a flat $SO(32) $ bundle
described by five commuting Wilson lines $W_j$.  (For the moment
we take the flat bundle to be topologically trivial, a point we return
to in section 3.4.)  
If one is free to vary all of these, one can certainly observe
the full $SL(5,\Z)$.  Suppose, though, that in some method of calculation,
the Wilson lines are frozen at particular values, and one can only
vary the metric and $B$-field.  Then one will only observe the subgroup
of $SL(5,\Z)$ that leaves the Wilson lines invariant.

For instance, if the Wilson lines are trivial -- a rather special
situation with unbroken $SO(32)$ -- one will see all of $SL(5,\Z)$.
Here is another case that will enter below though it will appear
mysterious at the moment.  As the $W_j$ commute, they
can be simultaneously diagonalized, with eigenvalues $\lambda_j^a$,
$a=1,\dots,32$.  Suppose that the $\lambda_j^a$ are all $\pm 1$, and
have the property that for each fixed $a$, $\prod_j\lambda_j^a=-1$.
There are 16 collections of five $\pm 1$'s whose product is $-1$
(namely $1,1,1,1,-1$ and four permutations of that sequence; 
$1,1,-1,-1,-1$ and nine
permutations of that sequence; 
and $-1,-1,-1,-1,-1$).  Let the $\lambda_j^a$ be such
that each such permutation appears exactly twice.  This breaks $SL(5,\Z)$
to the finite index  subgroup $\Gamma$ consisting of $SL(5,\Z)$ matrices
$M^j{}_k$, $j,k=1,\dots,5$ such that $\sum_jM^j{}_k$ is odd for each
fixed $k$.  If the Wilson lines are frozen at the stated values, it is
only $\Gamma$ and not all of $SL(5,\Z)$ that can be observed by
varying the metric and $B$-field.

\subsec{ Interpretation Of The Symmetry}

Let us go back to the Type IIB theory on K3 and the attempt
to interpret the strong coupling limit that was described,
the one with a visible $SL(5,\Z)$. 
As in the example just discussed,
the $SL(5,\Z)$ symmetry is strongly suggestive of the mapping class
group of a five-torus.   Thus,
one is inclined to relate this particular limit of Type IIB on K3
to the $M$-theory on $\R^6$ times a five-manifold built from $(\S^1)^5$.
This cannot be $(\S^1)^5$ itself, because the $M$-theory on
$\R^6\times ({\bf S}^1)^5$ would have twice as much supersymmetry as we want.
One is tempted instead to take an orbifold of $({\bf S}^1)^5$ in
such a way as to break half of the supersymmetry while preserving
the $SL(5,{\bf Z})$.

A natural way to break half the supersymmetries by orbifolding is to
divide by 
 a $\Z_2$ that acts as $-1$ on all five circles.
 This is actually the only choice that breaks half the supersymmetry
and gives a {\it chiral} $N=4$ supersymmetry in six dimensions.  In
fact, dividing by this $\Z_2$ 
leaves precisely those supersymmetries whose generators
obey $\Gamma^7\Gamma^8\dots \Gamma^{11}\epsilon=\epsilon$.  This condition
was encountered in the discussion of the five-brane, and leaves
the desired chiral supersymmetry.  So $M$-theory
compactified on $({\bf S}^1)^5/\Z_2$ is our candidate for an eleven-dimensional
interpretation of Type IIB superstrings on K3. 

More precisely, the proposal is that $M$-theory on $({\bf S}^1)^5/\Z_2$
has the property that when {\it any} ${\bf S}^1$ factor in
$({\bf S}^1)^5/\Z_2$ goes to zero radius, the $M$-theory on this
manifold goes over to a weakly coupled Type IIB superstring.
This assertion should hold not just for one of the five circles
in the definition of $(\S^1)^5/\Z_2$, but for any of infinitely many
circles obtained from these by a suitable symmetry transformation.

\subsec{ Anomalies}

Let us work out the massless states of the theory, first (as in
\ref\horava{P. Horava and E. Witten, ``Heterotic And Type I String
Dynamics From Eleven Dimensions,'' hepth/9510209.}) 
the ``untwisted states,'' that is the states that come directly
from massless eleven-dimensional fields, and then  the
``twisted states,'' that is, the states that in a macroscopic description
appear to be supported at the classical singularities of
$({\bf S^1})^5/\Z_2$.  

The spectrum of untwisted states can be analyzed  quickly by looking
at antisymmetric tensors.  The three-form $A$ of the eleven-dimensional
theory is odd under parity (because of the $A\wedge F\wedge F$ supergravity
interaction).  Since the $\Z_2$ by which we are dividing $({\bf S}^1)^5$
reverses orientation, $A$ is odd under this transformation.
The zero modes of $A$ on $({\bf S}^1)^5$ therefore give, after the
$\Z_2$ projection, five two-forms (and ten scalars, but no vectors
or three-forms) on $\R^6$.  The self-dual parts of these tensors are the
expected five self-dual tensors of the supergravity multiplet, 
and the anti-self-dual
parts are part of five tensor multiplets.  The number of tensor
multiplets from the untwisted sector is therefore five.

Just as in \horava, the untwisted spectrum is anomalous; there are
five tensor multiplets, while 21 would be needed to cancel the
gravitational anomalies.  16 additional tensor multiplets are
needed from twisted sectors.  

The problem, as independently raised in \mukhi, is that there appear
to be 32 identical twisted sectors, coming from the 32 fixed points
of the $\Z_2$ action on $(\S^1)^5$.  How can one get 16 tensor multiplets
from 32 fixed points?  We will have to abandon the idea of finding
a vacuum in which all fixed points enter symmetrically.

Even so,   there seems to be a paradox.
As explained in \horava, since the eleven-dimensional theory has no
gravitational anomaly on a smooth manifold,
the  gravitational anomaly of the eleven-dimensional massless fields
on an orbifold is a sum of delta functions supported at the fixed points.
In the case at hand, the anomaly can be canceled by 16 tensor multiplets
(plus a Green-Schwarz mechanism), but  there are 32 fixed points.
Thus, each fixed point has an anomaly, coming from the massless 
eleven-dimensional fields, that could be canceled by
$16/32=1/2$ tensor multiplets.\foot{The eleven-dimensional
massless fields by obvious symmetries contribute the same anomaly
at each fixed point.}  The paradox is that it is not
enough to {\it globally} cancel the gravitational anomaly
by adding sixteen tensor multiplets.  One needs to cancel the anomaly
{\it locally} in the eleven-dimensional world, somehow modifying
the theory to add at each fixed point half the anomaly of a tensor multiplet.
  How can this be, given that the
tensor multiplet is the only matter multiplet of chiral $N=4$ supersymmetry,
so that any matter system at a fixed point would be a (positive) integral
number of tensor multiplets?

\subsec{ Resolution Of The Paradox}

To resolve this paradox, the key point is that because the
fixed points in $({\bf S}^1)^5/\Z_2$ have codimension {\it five},
just like the codimension of a five-brane world-volume, there
is another way to cancel anomalies apart from including massless
fields on the world-volume.  We can assume that the fixed points
are magnetic sources of the four-form $F$.  In other words,
we suppose that (even in the absence of conventional five-branes)
$dF$ is a sum of delta functions supported at the orbifold fixed points.
If so, then  the bulk interaction $\Delta L=\int F\wedge I_7$ that
was discussed earlier will give additional contributions to the
anomalies supported on the fixed points.

Since a magnetic coupling of $F$ to the five-brane cancels the
anomaly of a tensor multiplet, if an orbifold fixed point has
``magnetic charge'' $-1/2$, this will cancel the anomaly from
the eleven-dimensional massless fields (which otherwise could be canceled
by $1/2$ a tensor multiplet).  If an orbifold fixed point has magnetic
charge $+1/2$, this doubles the anomaly, so that it can be canceled
if there is in addition a ``twisted sector'' tensor multiplet supported
at that fixed point.  Note that it is natural that a $\Z_2$ orbifold
point could have magnetic charge that is half-integral in units of the
usual quantum of charge.

A constraint comes from the fact that the sum of the magnetic charges
must vanish on the compact space $(\S^1)^5/\Z_2$.   Another
constraint comes from the fact that if we want to maintain supersymmetry,
the charge for any fixed point cannot be less than $-1/2$.  Indeed,
a fixed point of charge less than 
$-1/2$ would have an anomaly that could not be
canceled by tensor multiplets; a negative number of tensor multiplets
or a positive number of wrong chirality tensor multiplets (violating
supersymmetry) would be required.  
An example of how to satisfy these constraints and ensure local cancellation
of anomalies is to assign charge $-1/2$ to 16 of the fixed points,
and charge $+1/2$ to the other 16.  With one tensor multiplet supported
at each of the last 16 fixed points, such a configuration has
all anomalies locally cancelled in the eleven-dimensional sense.

Here is another way to cancel the anomalies locally.  Assign magnetic
charge $-1/2$ to each of the 32 fixed points, but include at each of 16 points
on $(\S^1)^5/\Z_2$ a conventional five-brane, of charge 1.  The total
magnetic charge vanishes (as $32(-1/2)+16=0$) and since both
a fixed point of charge $-1/2$ and a conventional five-brane are
anomaly-free, all anomalies are cancelled locally.  Each five-brane
supports one tensor multiplet; the scalars in the tensor multiplets
determine the positions of the five-branes on $(\S^1)^5/\Z_2$.

I would like to suggest that this last anomaly-canceling mechanism
is the general one,  and that
the case that the magnetic charge is all supported on the fixed points
is just a special case in which the five-branes and fixed points coincide.
In fact, if  a five-brane
happens to move around and meet a fixed point, the charge of that
fixed point increases by 1.  This gives a very natural interpretation of
the ``twisted sector'' modes of a fixed point of charge 1/2.  Such  a fixed
point supports a tensor multiplet, which contains five scalars; we 
interpret the scalars as representing a possible perturbation in the
five-brane position away from the fixed point.

If we accept this interpretation, there is no issue of what is the
``right'' configuration of charges for the fixed points; 
any configuration obeying the 
constraints (total charge 0 and charge at least $-1/2$ for each fixed point)
appears somewhere on the moduli space.  The only issue is what
configuration of charges has the most transparent relation to string theory.

Let us parametrize the five circles in $(\S^1)^5$ by periodic
variables $x^j,\,j=7,\dots,11$, of period 1, with $\Z_2$ acting
by $x^j\to -x^j$ so that the fixed points have all coordinates
$0$ or $1/2$.  We take $SL(5,\Z)$ to act linearly, by $x^j\to M^j{}_kx^k$,
with $M^j{}_k$ an $SL(5,\Z)$ matrix.  Thus $SL(5,\Z)$ leaves invariant
one fixed
point $P$, the ``origin'' $x^j=0$, and acts transitively on the other 31.
The only $SL(5,\Z)$-invariant configuration of charges obeying
the constraints is to assign
magnetic charge $+31/2$ to $P$ and $-1/2$ to each of the others.
Then each of the 16 tensor multiplets would be supported at the origin.
This configuration cancels the anomalies and is $SL(5,\Z)$ invariant.
However, it does not seem to be the configuration with the closest
relation to string theory.

To see this, consider the limit in which one of the circles in $(\S^1)^5$
becomes small.  To an observer who does not detect this circle,
one is then left with $(\S^1)^4/\Z_2$, which is a K3 orbifold.  Our
hypothesis about  $M$-theory on      $(\S^1)^5/\Z_2$ 
says that this theory should go over to weakly coupled Type IIB on
K3 when {\it any} circle shrinks.  In $(\S^1)^4/\Z_2$, there are
16 fixed points; in quantization of Type IIB superstrings on this
orbifold, one tensor multiplet comes from each of the   16 fixed points.

In $M$-theory on $(\S^1)^5/\Z_2$, there are   32 fixed points.
When one of the circles is small, then -- to an observer who does not
resolve that circle -- the 32 fixed points appear to coalesce pairwise
to the 16 fixed points of the string theory on $(\S^1)^4/\Z_2$.
To reproduce the string theory answer that one tensor multiplet comes
from each singularity, we want to arrange the charges on $(\S^1)^5/\Z_2$
in such a way that each pair  of fixed points differing only in the values
of one of the coordinates contributes one tensor multiplet.

This can be done by arranging the charges in the following ``checkerboard''
configuration.  If a fixed point has $\sum_j x^j$ integral, we give
it charge $-1/2$.  If $\sum_jx^j$ is a half-integer, we give it charge
$+1/2$.  Then any 
two fixed points differing only by the value of the $x^j$ coordinate
 -- for any given $j$ -- have equal and opposite charge, and contribute
a total of one tensor multiplet. 

Moreover, the four-form field strength $F$ of the $M$-theory reduces
in ten dimensions to a three-form field strength $H$.  This vanishes
for string theory on K3, so one can ask how the string theory can be
a limit of an eleven-dimensional theory in a vacuum with non-zero $F$.
If we arrange the charges in the checkerboard fashion, this puzzle
has a natural answer.  In the limit in which the $j^{th}$ circle
shrinks to zero, equal and opposite charges are superposed and cancel,
so the resulting ten-dimensional theory has zero $H$.

The checkerboard configuration is not invariant under all of $SL(5,\Z)$,
but only under the finite index  subgroup $\Gamma$ introduced just prior
to section 3.1  (the subgroup consisting of matrices $M^j{}_k$ such
that $\sum_j M^j{}_k$ is odd for each $k$).  Thus the reduction
to ten-dimensional string theory can work not only if one shrinks
one of the five circles in the definition of $(\S^1)^5/\Z_2$, but also
if one shrinks any of the (infinitely many) circles obtainable
from these by a $\Gamma$ transformation.

Just as in the discussion in which $\Gamma$ was introduced,
 in the checkerboard
vacuum, one cannot see the full $SL(5,\Z)$ if the only parameters
one is free to vary are the metric and three-form $A$ on $(\S^1)^5/\Z_2$.
An $SL(5,\Z)$ transformation $w$ not in $\Gamma$ is a symmetry only
if combined with a motion of the other moduli -- in fact a motion of
some five-branes to compensate for the action of $w$ on the charges
of fixed points.  

\subsec{Check By Comparison To Other Dualities}

In the study of string theory dualities,
once a conjecture is formulated that runs into
no immediate contradiction, one of the main ways to test it is to 
try to see what implications it has when combined with other,
better established dualities.

In the case at hand, we will (as was done independently by Dasgupta
and Mukhi \mukhi) mainly compare our hypothesis about $M$-theory
on   $(\bf S^1)^5/\Z_2$ to the assertion that $M$-theory on 
$X\times \S^1$ is equivalent, for any five-manifold $X$, to Type IIA
on $X$.

To combine the two assertions in an interesting way, we consider
$M$-theory on $\R^5\times \S^1\times (\S^1)^5/\Z_2$.  On the one
hand, because of the $\S^1$ factor, this should be equivalent
to Type IIA on $\R^5\times (\S^1)^5/\Z_2$, and on the other
hand, because of the $(\S^1)^5/\Z_2$ factor, it should be equivalent
to Type IIB on $\R^5\times \S^1\times {\rm K3}$.

It is easy to see that, at least in general terms, we land on our
feet.  Type IIB on $\R^5\times \S^1\times {\rm K3}$ is equivalent
by $T$-duality to Type IIA on $\R^5\times \S^1\times {\rm K3}$,
and the latter is equivalent by Type IIA - heterotic duality
to the heterotic string on $\R^5\times \S^1\times \T^4=\R^5\times \T^5$,
and thence by heterotic - Type I duality to Type I on
$\R^5\times\T^5$.  

\nref\polchinski{J. Dai, R. Leigh, and J. Polchinski, ``New Connections
Between String Theories,'' Mod. Phys. Lett. {\bf A4} (1989) 2073.}
\nref\oldhorava{P. Horava, ``Strings On World Sheet Orbifolds,'' Nucl.
 Phys. {\bf B327} (1989) 461, ``Background Duality Of Open String
Models,'' Phys. Lett. {\bf B321} (1989) 251.}
On the other hand, Type IIA on $\R^5\times (\S^1)^5/\Z_2$
is an orientifold which is  equivalent by $T$-duality to
Type I on $\R^5\times \T^5$ \refs{\polchinski,\oldhorava}.

So the prediction from our hypothesis about $M$-theory on $(\S^1)^5/\Z_2$
-- that Type IIA on $(\S^1)^5/\Z_2$ should be equivalent to Type IIB
on $\S^1\times {\rm K3}$ -- is correct.
This is a powerful test.
 
\bigskip
\noindent{\it Components Of The Moduli Space}

What remains to be said?  The strangest part of our discussion
about $M$-theory on $(\S^1)^5/\Z_2$ was the absence of a vacuum
with symmetry among the fixed points.
We would like to find a counterpart of this at the string theory level,
for Type IIA on $(\S^1)^5/\Z_2$.

The Type IIA orientifold on $(\S^1)^5/\Z_2$ needs -- to cancel anomalies
-- 32 $D$-branes  
located at   32 points  in $(\S^1)^5$; moreover, this configuration
of 32 points must be invariant under $\Z_2$.  
It is perfectly possible to place one $D$-brane at each of the 32 fixed
points, maintaining the symmetry between them.  Does this not contradict
what we found in eleven dimensions?

The resolution of this puzzle starts by observing that the $D$-branes
that are not at fixed points are paired by the $\Z_2$.  So as the $D$-branes
move around in a $\Z_2$-invariant fashion, the number of $D$-branes
at each fixed point is conserved modulo two (if a $D$-brane approaches
a fixed point, its mirror image does also).  Thus, there is a $\Z_2$-valued
 invariant associated with each fixed point; allowing
for the fact that the total number of $ D$-branes is even, there are
31 independent $\Z_2$'s.

What does this correspond to on the Type I side?  A configuration of
32 $ D$-branes on $(\S^1)^5/\Z_2$ is $T$-dual to a Type I theory
compactified on $(\S^1)^5$ with a flat $SO(32)$ bundle.  
However, the moduli space of flat $SO(32)$ connections on the five-torus
is not connected -- there are many components.  One component contains
the trivial connection and leads when one considers the deformations
to the familiar Narain moduli space of the heterotic string on the five-torus.
This actually corresponds to a $D$-brane configuration with an even
number of $D$-branes at each fixed point.  The Wilson lines $W_j$ can
be simultaneously block-diagonalized, with 16 two-by-two blocks.
The $a^{th}$ block in $W_j$ is
\eqn\rufu{\left(\matrix{ \cos \theta_{j,a} & \sin\theta_{j,a} \cr
                        -\sin\theta_{j,a}  & \cos\theta_{j,a} \cr}\right),}
with $\theta_{j,a}$, $j=1,\dots,5$ being angular variables that
determine the position on $(\S^1)^5$ of the $a^{th}$ $D$-brane
(which also has an image whose coordinates are $-\theta_{j,a}$).

There are  many other components of the moduli space of flat
connections on the five-torus, corresponding to the 32 $\Z_2$'s noted
above.  Another component -- in a sense at
the opposite extreme from the component that contains the trivial connection
 -- is the following.  Consider
a flat connection with the properties that the $W_j$ can all
be simultaneously diagonalized, with  eigenvalues $\lambda_{j,a}=\pm 1$, 
$a=1,\dots,
32$.   Since the positions of the $D$-branes are the phases of the
eigenvalues of the $W_j$, 
this corresponds to a situation in which all $D$-branes are at fixed points.
Pick the $\lambda_{j,a}$ such that each of the 32 possible sequences
of five $\pm 1$'s arises as $\lambda_{j,a}$ for some value of $a$.  Then
there is precisely one $D$-brane at each of the 32 fixed points.  This
flat bundle -- call if $F$ -- cannot be deformed {\it as a flat bundle}
to the flat bundle with trivial connection; that is clear from
the fact that the number of $D$-branes at fixed points is odd.  Therefore,
$F$ does not appear on the usual  Narain moduli space of toroidal
compactification of the heterotic string to five dimensions.  However, 
it can be shown that the bundle $F$ is topologically trivial so that
the flat connection on it can be deformed (but not via flat connections)
to the trivial connection.\foot{In a previous draft of this paper,
it was erroneously claimed that the bundle $F$ was topologically non-trivial,
with non-vanishing Stieffel-Whitney classes.  The error was pointed
out by E. Sharpe and some topological details were clarified by D. Freed.}  
Thus, compactification using the bundle
$F$ is continuously connected to the usual toroidal compactification,
but only by going through configurations that are not classical solutions.

The fact that the configuration with one $D $-brane at each fixed 
point is not on the usual component of the moduli space
 leads to a solution to our puzzle.  In reconciling the
two string theory descriptions of $M$-theory                   on 
$\R^5\times \S^1\times (\S^1)^5/\Z_2$, a key step was Type IIA - heterotic
string duality relating Type IIA on $\R^5\times \S^1\times {\rm K3}$ to the
heterotic string on $\R^6\times \S^1\times( \S^1)^4=\R^5\times \T^5$.  
This    duality
holds with the {\it standard} component of the moduli space on
 $\T^5$, so even though
the symmetrical $D$-brane configuration exists, it is not relevant
to our problem because it is related to a different component of the 
moduli space of flat $SO(32)$ bundles.

Working on the Type IIA orientifold on $(\S^1)^5/\Z_2$ which is $T$-dual
to a flat $SO(32)$ bundle on the usual component of the moduli space
means that the number of $D$-branes at each fixed point is even.  
With 32 $D$-branes and 32 fixed points, it is then impossible to
treat symmetrically all fixed points.  One can, however, pick any    16
fixed points, and place two $D$-branes at each of those, and none at the 
others.  In the quantization, one then gets one five-dimensional
vector multiplet from each     fixed point that is endowed with
a $D$-brane and none from the others.\foot{This is most easily seen
by perturbing to a situation in which the pair of $D$-branes is near
but not at the fixed point.  For orientifolds, there are no twisted
sector states from a fixed point that does not have $D$-branes.
After the $\Z_2$ projection, a pair of $D$-branes in the orientifold
produces the same spectrum as a single  $D$-brane in an unorientifolded
Type IIA, and this is a single vector multiplet, as explained
in detail in section 2 of 
\ref\ugwitten{E. Witten, ``Bound States Of Strings
And $p$-Branes,'' hepth/9510135.}.}
  Recalling that the vector
multiplet is the dimensional reduction of the tensor multiplet
from six to five dimensions, this result agrees with what we had in
eleven dimensions: given any 16 of the      32 fixed points, there
is a point in moduli space such that each of those 16 contributes
precisely one matter multiplet, and the others contribute none.

It is possible that the absence of a vacuum with symmetrical treatment
of all fixed points means that these theories cannot be strictly
understood as orbifolds, but in any event, whatever the appropriate
description is in eleven dimensional $M$-theory, we have found
a precisely analogous behavior in the ten-dimensional Type IIA orientifold.

\bigskip
\noindent{\it Other Similar Checks}

One might wonder about other similar checks of the claim about
$M$-theory on $(\S^1)^5/\Z_2$.  One idea is to look at
$M$-theory on $ \R^5\times \S^1/\Z_2\times (\S^1)^5/\Z_2$.
The idea would be that this  should turn into an $E_8\times E_8$
 heterotic string upon taking the $\S^1/\Z_2$
small, and into a Type IIB orientifold on $\S^1/\Z_2\times K3$ 
if one shrinks the $(\S^1)^5/\Z_2$.  However, because the      two 
$\Z_2$'s do not commute in acting on spinors, it is hard to make
any sense of this orbifold.

A similar idea is to look at $M$-theory on $\R^4\times {\rm K3}
\times ({\bf S}^1)^5/\Z_2$.  When the last factor shrinks, this should
become Type IIB on ${\rm K3}\times {\rm K3}$, while if the ${\rm K3}$ 
factor shrinks then (allowing, as in a discussion that will appear
elsewhere \ref\duffman{M. Duff, R. Minasian, and E. Witten, ``Evidence
For Heterotic/Heterotic Duality,'' to appear.}, 
for how the $\Z_2$ orbifolding acts on the homology of K3)
one gets the         heterotic string on $(\S^1)^8/\Z_2$.  These
should therefore be equivalent.  But one does not immediately
have tools to verify or disprove that equivalence.

\bigskip
\noindent{\it Relation To Extended Gauge Symmetry And Non-Critical
Strings}

A rather different kind of check can be made by looking at the behavior
when some $D$-branes -- or eleven-dimensional five-branes -- coincide.

Type IIA on K3 gets an extended $SU(2)$ gauge symmetry when the K3
develops an $A_1$ singularity.\foot{We really mean a quantum
$A_1$ singularity including a condition on a certain world-sheet
theta angle \ref\aspinwall{P. Aspinwall, ``Extended Gauge Symmetries And
K3 Surfaces,'' hep-th/9507012.}.}  This is not possible for Type IIB
on K3, which has a chiral $N=4$ supersymmetry that forbids vector
multiplets.
  Rather, the weakly coupled
Type IIB theory on a K3 that is developing an $A_1$ singularity
 develops \ref\witcom{E. Witten, ``Some Comments On String Dynamics,''
hep-th/9507121.}
a non-critical string (that is, a string that
propagates in flat Minkowski space and does not have the graviton as
one of its modes) that couples to the anti-self-dual part of one of
the antisymmetric
tensor fields (the part that is in a tensor multiplet, not in the supergravity
multiplet).

This six-dimensional non-critical string theory is a perhaps rather
surprising  example, apparently, of a non-trivial 
quantum theory in six-dimensional Minkowski space.  Recently, it
was argued by Strominger
\ref\stromnew{A. Strominger, ``Open $P$-Branes,'' hep-th/9512059.}
that by  considering almost coincident parallel five-branes
in eleven dimensions, one gets on the world-volume an alternative
realization of the same six-dimensional non-critical string theory.

We can now (as  partly anticipated by Strominger's remarks)
close the circle and deduce from the relation between  $M$-theory
on $\T^5/\Z_2$ and Type IIB on K3
{\it why} Type IIB on a K3 with an $A_1$ singularity
gives the same unusual low energy dynamics as two nearby parallel
five-branes in eleven dimensions.  This follows from the fact
that  in the map from $M$-theory
on $\T^5/\Z_2$ to Type IIB on K3, a configuration on $\T^5/\Z_2$ with
two coincident five-branes is mapped to a K3 with an $A_1$ singularity.
To see that these configurations are mapped to each other, it is enough
to note that upon compactification on an extra circle of generic radius, they 
are   precisely the configurations that
 give an enhanced $SU(2)$.  This may be deduced
as follows:

(1)  $M$-theory on $\R^5\times \S^1\times \T^5/\Z_2$
is equivalent to Type IIA on $\R^5\times \T^5/\Z_2$, with the five-branes
replaced by $D$-branes, and gets an enhanced $SU(2)$ gauge symmetry precisely
when
two   five-branes, or $D$-branes, meet.  Indeed, when two $D$-branes
meet, their $U(1)\times U(1)$ gauge symmetry (a $U(1)$ for each $D$-brane)
is enhanced to $U(2)$ (from the Chan-Paton factors of two coincident
$D$-branes), or equivalently a $U(1)$ is enhanced to $SU(2)$.

(2) Type IIB on $\R^5\times \S^1\times {\rm K3}$ is equivalent
to Type IIA on $\R^5\times \S^1\times {\rm K3}$ and therefore -- because
of the behavior of Type IIA on K3 -- the condition on the K3 moduli
that causes a $U(1)$ to be extended to $SU(2)$ is precisely that there should 
be
an $A_1$ singularity.

\bigskip
\noindent{\it Other Orbifolds}

Dasgupta and    Mukhi also discussed $M$-theory orbifolds
$\R^{11-n}\times (\S^1)^n/\Z_2$.  The $\Z_2$ action on the fermions
multiplies them by the matrix $\tilde \Gamma=
\Gamma^{11-n+1}\Gamma^{11-n+2}\dots \Gamma^{11}$, and the orbifold
can therefore only be defined if $\tilde \Gamma^2=1 $ (and not $-1$),
which restricts us to $n$ congruent to 0 or 1 modulo 4.

The case
$n=1$ was discussed in \horava, $n=4$ gives a K3 orbifold, and $n=5$
has been the subject of the present paper.  The next cases are $n=8,9$.
For $n=8$, as there are no anomalies, it would take a different approach to 
learn about the massless states from fixed points.  For $n=9$,
Dasgupta and Mukhi pointed out the beautiful fact that the number of
fixed points -- $2^9=512$ --
equals the number of left-moving massless fermions needed to cancel
anomalies, and suggested  that one such fermion comes from
each fixed point.  Since the left-moving fermions are singlets
under the (chiral, right-moving) supersymmetry, this scenario
is entirely compatible with the supersymmetry and is very plausible.

\bigskip
\noindent{\it Reduced Rank}

Finally, let us note the following interesting application
of part of the discussion above.  Toroidal compactification of the heterotic
(or Type I) string on a flat  $SO(32)$ bundle 
that is not on the usual component of the moduli space (being $T$-dual
to a configuration with an odd number of $D$-branes at fixed points) 
gives an interesting and simple way to reduce the rank
of the gauge group while maintaining the full supersymmetry. Since
$2n+1$ $D$-branes at a fixed point gives gauge group $SO(2n+1)$, one
can in this way get gauge groups that are not simply laced.
Models with these properties have been constructed via free
fermions \ref\chaudhuri{S. Chaudhuri, G. Hockney, and J. D. Lykken,
``Maximally Supersymmetric String Theories in $D<10$,'' hepth/9505054.}
and as   asymmetric orbifolds \ref\chpol{S. Chaudhuri and J.
Polchinski, ``Moduli Space Of CHL Strings,'' hepth/9506048.}.

\bigskip

I would like to thank M. Duff, J. Polchinski, and
 C. Vafa for helpful discussions.
\listrefs
\end